1

# DESIGN PRINCIPLES AND CLINICIAN PREFERENCES FOR PHARMACOGENOMIC CLINICAL DECISION SUPPORT ALERTS

Timothy M. Herr, PhD[1]; Therese A. Nelson, AM, LSW[2]; Justin B. Starren, MD, PhD, FACMI[1]

[1]Department of Preventive Medicine,

Northwestern University Feinberg School of Medicine, Chicago, IL

[2]NUCATS Institute,

Northwestern University Feinberg School of Medicine, Chicago, IL

**Abstract**

OBJECTIVE: To better understand clinician needs and preferences for the display of pharmacogenomic (PGx) information in clinical decision support (CDS) tools.

MATERIALS AND METHODS: We developed a semi-structured interview to collect feedback and preferences in six key areas of PGx CDS design, from clinicians who had prior experience with live PGx CDS tools. Eight clinicians from Northwestern Medicine's (NM) General Internal Medicine clinic participated in the study.

RESULTS: Clinicians expressed a preference for interruptive pop-up alerts during order entry that contain brief descriptions of relevant drug-gene interactions, with a clear and specific recommended alternative course of action when a medication is contraindicated. They did not wish to see detailed genetic data, instead preferring phenotypic information predicted from the genotype. Nor did they wish to be interrupted when genetic test results do not indicate a change in the treatment plan. Clinicians reported little familiarity with Clinical Pharmacogenetic

Implementation Consortium (CPIC) prescribing recommendations but reported trusting recommendations of their professional societies and resources like UpToDate. Analysis of unstructured comments concurred with structured results, indicating a general uncertainty among participants around how to interpret and apply PGx information in practice.

DISCUSSION: Results point to several underlying principles that can inform future PGx CDS alert designs, including: Be Specific and Actionable; Be Brief; Display Phenotypes not Genotypes; Rely on Sources Clinicians Already Trust; and, Be Adaptable to Learning Effects.

CONCLUSION: This study is part of a broader socio-technical design approach to PGx CDS design underway at NM and provides a baseline for future PGx CDS development. Designs based on these results have the potential to improve clinician education and adherence levels, and to improve patient outcomes.

**Background and Significance**

Genomic medicine continues to show promise for improving patient outcomes in myriad ways, with new discoveries of links between genetics and health regularly appearing in the scientific literature. Certain specific genetic tests have garnered broad public awareness (e.g. the BRCA genes[1]) and direct-to-consumer genetic testing is reaching the mainstream.[2] Notwithstanding, clinician comfort levels with genetic data and the use of genetic testing remains relatively low.[3] [4] Integration of genetic- and genomic-based factors into standard clinical practices must be done in a way that benefits both the clinician and the patient, allowing the best course of action to be followed with minimal burden and confusion.

Pharmacogenomics (PGx) is among the most promising areas of genomic medicine. It is regularly cited as a means to reduce adverse drug events, avoid ineffective drugs, and optimize drug and dose selection more reliably and with less trial-and-error.[5,6] Clinical recommendations for the use of PGx in drug prescribing exist but are not yet widely implemented.[7] Clinical decision support (CDS) that delivers PGx information and prescribing recommendations at the point of care is one promising way of disseminating this knowledge and improving patient care.[8-11]

Some PGx CDS implementations have been described in the literature. Most sites in the Electronic Medical Records and Genomics (eMERGE) Network have implemented PGx CDS systems.[12,13] Several other organizations have also succeeded in implementing first-generation PGx CDS systems, both in the United States[14-16] and other countries.[17] Institutions implementing PGx CDS tools have taken widely varying approaches – the University of Chicago's Genomic Prescribing System uses a "traffic light" model in a standalone system,[18,19] while St. Jude's PG4KDS system uses traditional pop-up alerts integrated into the EHR.[20] Even within the eMERGE network, where institutions collaborate on genomic medicine projects, significant variation in PGx CDS alert design made it difficult to aggregate results for cross-site analyses.[13] There is little information regarding which of these designs may be the most effective at conveying PGx knowledge to providers and improving care, and similarly, little consensus around optimal PGx CDS design.

Some studies have begun to examine PGx CDS alerts in an effort to optimize their designs, but this work has been too preliminary to drive a consensus on best practices for incorporating PGx knowledge into clinician workflows, or for displaying it in a way that is easy

4to understand and use by clinicians.[13,20-22] Due to the complexities of genomic medicine, and low levels of clinician education and comfort in the field, simpler approaches to PGx CDS development—such as design-by-committee processes—may not be effective.[23] Instead, more formal socio-technical design approaches that iterate on designs, deeply incorporate user workflow needs, and respond to user feedback at multiple phases of development may be more effective in creating genuinely useful PGx CDS tools.

**Objective**

This study aimed to apply socio-technical design methods to determine actual clinician preferences for PGx information at Northwestern Medicine, as a first step towards designing better-optimized PGx CDS alerts. This was accomplished through a series of semi-structured interviews conducted with clinicians who have real-world experience with a live PGx CDS system. The results of this study can be used to inform future improvements on PGx CDS designs. This study emphasized matching the user's workflow and information needs through a focus on appropriate timing and location for PGx data display, type of information conveyed, level of detail, specificity of recommendations, and trusted sources of PGx recommendations.

**Materials and Methods**

We developed a semi-structured interview that focused on areas of interest we identified after previous phases of NU's eMERGE-PGx project.[11] We interviewed eight clinicians from Northwestern Medicine's General Internal Medicine (GIM) outpatient clinic. Each of the participants had previous experience with our existing PGx CDS tools. Existing PGx CDS tools were built using the Best Practice Advisories functionality of the clinic's EHR (Epic Systems Corp., Verona, WI), based on genomic indicators received from an ancillary genomic system.[24] As described in detail below, interviews were transcribed and systematically analyzed to identify themes and attitudes that could inform the design of improved alerts. The Northwestern IRB approved all aspects of this study.

Building upon our previous research,[23] we identified six areas of interest for which we believed a better understanding could lead to improved alert designs and more positive clinician interactions. We developed an interview guide with a series of structured and open-ended questions to probe these areas. The areas of interest and methods for assessing them are listed in Table 1. All interviews began with a series of questions assessing basic demographics and PGx comfort level. Participants were provided the opportunity to elaborate upon their answers for all structured questions, if desired. The complete interview guide is available in Supplement 1.

Table 1 - Areas of Interest and Methods of Assessment

| Area of Interest | Method of Assessment |
|---|---|
| Understanding Workflow | 1 open-ended question (based on a model workflow diagram) |
| Feedback on Existing Tools | 3 Yes/No questions<br>1 open-ended question |
| Appropriate Locations for PGx Information | 7 Likert-scale questions (7-point scale) |
| Level of Detail and Interruption for Alerts | 4 Preference-ranked questions (4 options) |
| Types of Information to Display in Alerts | 13 Likert-scale questions (7-point scale) |
| Trusted Knowledge Sources for PGx | 3 Yes/No questions<br>3 open-ended questions |

Because we were interested in receiving feedback on the workflows and alerts used in the existing PGx CDS tools in place at NM, we limited recruitment to clinicians who had previously interacted with those tools in a real-world clinical setting. Through a query of EHR data available in the Northwestern Medicine Enterprise Data Warehouse (NMEDW), we identified all clinicians who had previously recorded an interaction with at least two PGx CDS alerts, were still NM employees at the time of the study, and were an MD, PA, or LPN. Based on these criteria, 27 qualifying clinicians were identified and recruited for participation. Each qualifying clinician was contacted by e-mail, asked to participate in a 30-minute interview, and promised a $25 gift card for their time. Eight clinicians agreed to participate. All interviews were conducted in person, either at the clinician's office or in a suitable conference room. Interview sessions were recorded and later transcribed by a third-party transcription service. Each transcript was reviewed and edited for accuracy by TMH.

Analysis was divided according to the structured or open-ended nature of the responses. All structured responses were reviewed and aggregated directly by the first author. Unstructured responses (either responses to open-ended questions or elaborations upon structured responses) were analyzed by two independent coders to identify themes and attitudes that were not captured by the structured responses. TMH performed the first coding pass and created a codebook with 97 individual codes roughly organized in parallel with the interview guide. This codebook, and un-coded copies of the transcripts, was provided to TAN, who performed a second coding pass without knowledge of TMH's prior coding decisions. TMH and TAN met in a series of sessions to compare coding choices and reconcile any discrepancies via one-on-one discussion. Through these meetings, 100% consensus was reached without need for third-party intervention.

To identify the themes and axes that emerged from the coding process, we performed a two-pass analysis of the codebook. TMH created physical cards individually printed with each code and description from the codebook and performed a card sort by grouping codes into related groups. TAN and JBS collaboratively performed a second card sort, without knowledge of TMH's decisions. The codes and categories for each pass were entered into a spreadsheet. All three authors jointly reviewed the results of the two passes and reconciled differences to finalize the overall themes used for the unstructured data analysis.

**Results**

Basic Demographics

The eight participants averaged 21.5 years of clinical experience. Seven (87.5%) were board certified in only internal medicine. One participant (12.5%) was certified in both internal medicine and cardiology. On average, participants were scheduled to see patients 17 hours per week.

Understanding Workflow

When presented with our model diagram representing a typical prescribing workflow, seven of the eight participants (87.5%) reported that they review medications earlier in their workflow than in the diagram. There were no other major discrepancies between our diagram and actual clinician workflow. The updated version, representing typical clinician workflow as reported by the participants, is presented in Figure 1.

Figure 1 - Archetypal GIM Clinician Prescribing Workflow at NM

*A consensus, idealized prescribing workflow, as agreed upon by general internal medicine physicians at Northwestern Medicine. Real-world processes may include loop-backs and other variations.*

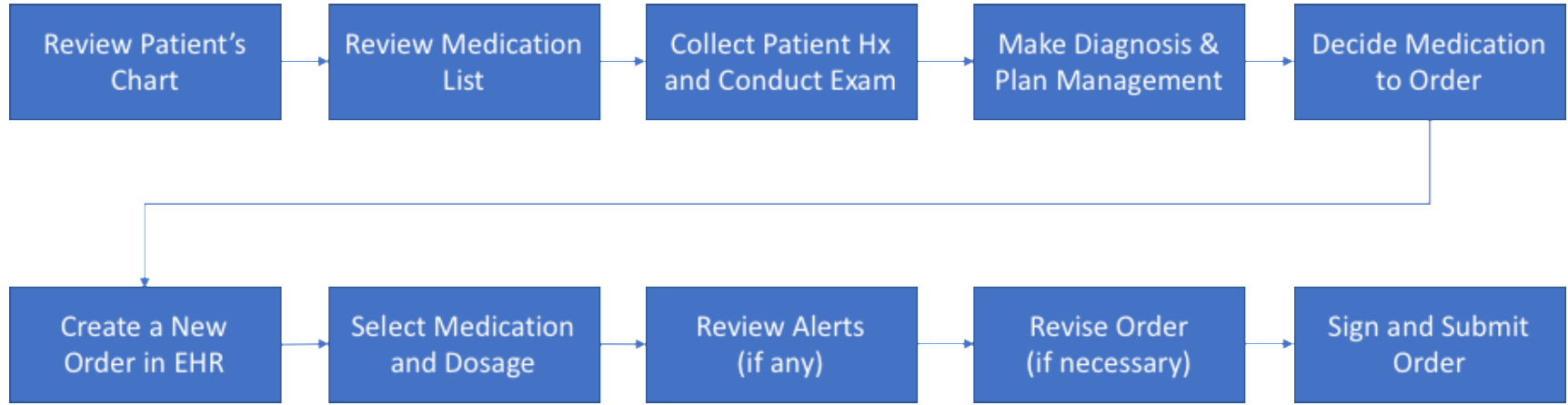

Feedback on Existing Tools

When asked about their thoughts on the existing PGx CDS tools, participants indicated a general ambivalence towards the tools, with clear room for improvement and no clear hostility towards the use of PGx CDS. All eight participants (100%) reported that they recalled seeing PGx CDS alerts at some point in their practice. When asked whether they felt the alerts were useful for them in making a prescribing decision, responses were overall neutral, with four (50%) finding them helpful and four (50%) not finding them helpful. Six (75%) of the physicians had suggestions for improvements and two (25%) did not.

Participants also made suggestions for improvement. The most common suggestions were related to making it easier to act on the alerts. Possible improvements included providing clearer guidance and specific alternative recommended medications. This topic is explored in additional depth in the results of the unstructured data analysis, below.

When asked to speculate on why real-world PGx CDS response rates for the existing tools are relatively low (see previous publications[13,23]), the most common response was that the recipient of the alert may not be the right target audience. Six participants (75%) reported some variation on this idea; for instance, a different specialty may be more appropriate, or they may not have been the one to initiate the drug and do not want to modify it. Five (62.5%) believed that it could be due to patients that are already successfully on a medication and a lack of desire to change what is working.

Appropriate Locations for PGx Information

When asked about potential locations for presenting PGx information in the EHR, two of the seven options had zero participants rate them negatively on the seven-point Likert scale (ranging from Must Avoid to Must Have). "As an alert during order entry, when relevant" was the most approved-of option, with five participants (62.5%) rating it Very Helpful and three (37.5%) rating it a Must Have. "As an entry on the Problem List" also had no negative responses, with five participants rating it at least Helpful and three rating it Neutral. All response rates are shown in Figure 2.

Figure 2 – Preferences for PGx Information Location in the EHR

*Number of participants expressing level of helpfulness of seven different EHR locations for pharmacogenomic information.*

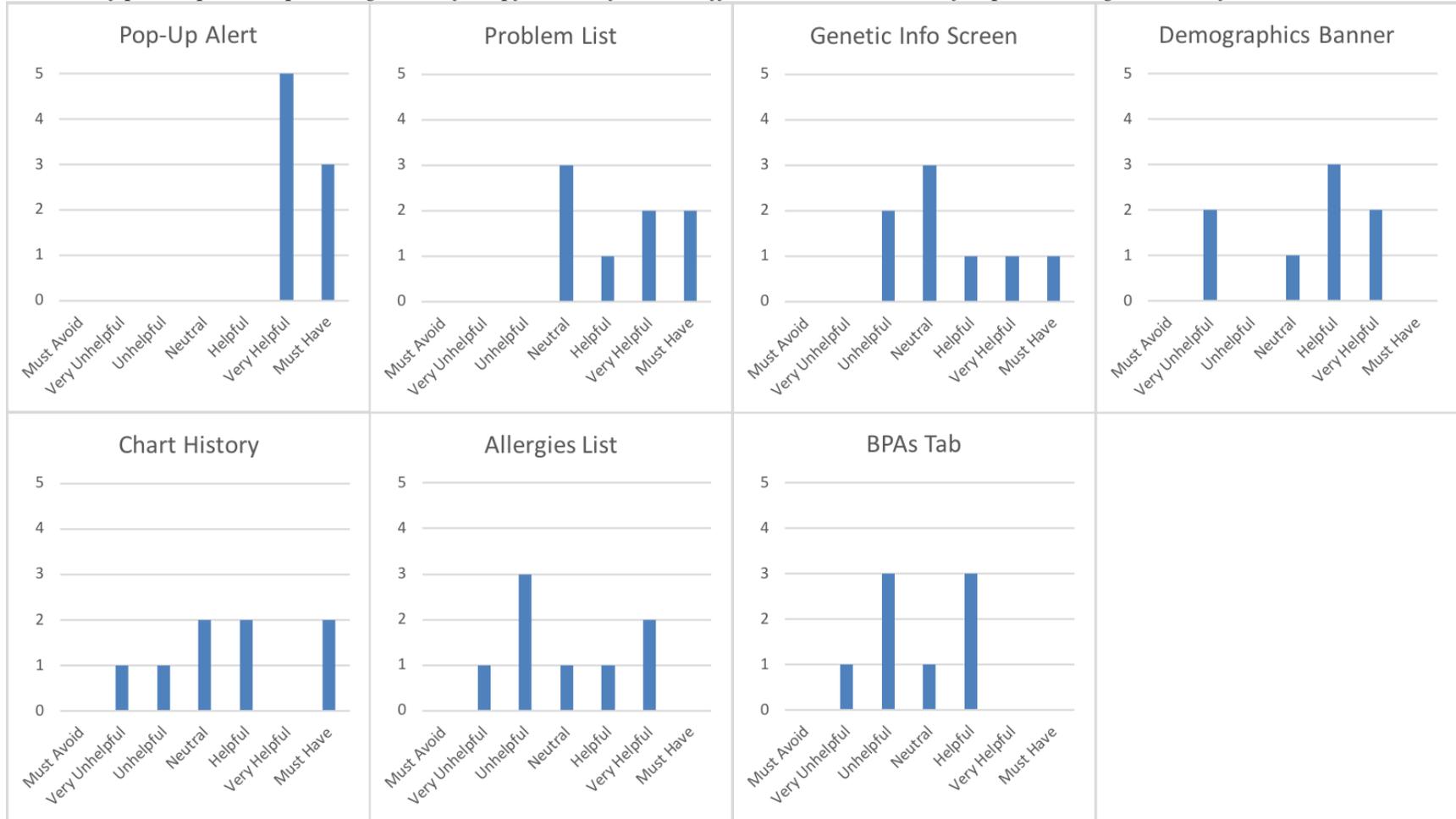

Level of Detail and Interruption for Alerts

When asked about the level of detail and interruption an alert should provide, there was a general trend towards a preference for brevity in the information presented. Additionally, clinicians did not want to be interrupted for alerts that were primarily informational in nature, although they did prefer interruption when an action was required. The overall ranking for each display option was determined by the mode value. Each queried scenario and the aggregated clinician-reported rankings are listed in Table 2.

Table 2 – Preferred Level of Detail and Interruption for Alerts, by Mode Rankings

*Consensus user rankings from 1 (most preferred) through 4 (least preferred) of four different options for level of detail and interruption in a Pharmacogenomic Clinical Decision Support alert, as determined by Mode ranking; "Detailed Alert" and "Brief Alert w/ Links" are fully interruptive options*

| Scenario | Level of Detail and Interruption | | | |
|---|---|---|---|---|
| | **Detailed Pop-up Alert** | **Brief Pop-up Alert w/ Links** | **Non-Interruptive Alert** | **No Alert** |
| Your patient has relevant genetic data on file, and that data <u>DOES</u> indicate a change from the medication or dosage you are selecting. | 2 | **<u>1</u>** | 3 | 4 |
| Your patient has relevant genetic data on file, but that data <u>DOES NOT</u> indicate a change from the medication or dosage you are selecting. | 4 | 3 | 2 | **<u>1</u>** |
| Your patient has relevant genetic data on file, but that data includes a <u>Variant of Unknown Significance</u> with no clear clinical impact. | 4 | 3 | 2 | **<u>1</u>** |
| Your patient <u>DOES NOT</u> have relevant genetic data on file, but the medication or dosage you are selecting has known genetic interactions. | 2 | **<u>1</u>** | 3 | 4 |

Types of Information to Display in Alerts

When asked about what type of information they prefer to see in a PGx CDS alert when a patient's genetic profile indicates a medication change, participants clearly preferred two of the options. The most approved-of options were "A specific recommended alternative medication and/or dosage," with five participants (62.5%) rating it as a Must Have, two (25%) rating it as Very Helpful, and one (12.5%) rating it as Helpful and "A clinically-relevant phenotypic interpretation of the genetic test results (e.g., Clopidogrel Intermediate Metabolizer)," with three participants (37.5%) rating it as a Must Have and five (62.5%) rating it as Very Helpful. Conversely, "A complete, detailed description of the drug-gene interaction, and how prescribing is affected, taken from the academic literature" was not well-received, with three participants (37.5%) rating it as Very Unhelpful, three (37.5%) rating it as Neutral, and two (25%) rating it as Helpful. The option "Your patient's actual genetic data in the format returned by the testing laboratory (e.g., CYP2C19 *1/*17)," was also rated negatively, with two participants (25%) rating it as Very Unhelpful, one (12.5%) rating it as Unhelpful, two (25%) rating it as Neutral, and three (37.5%) rating it as Helpful. All response rates are shown in Figure 3.

Figure 3 – Preferences for Types of PGx Information in Alerts

*Number of participants expressing level of helpfulness of six different types of pharmacogenomic information in clinical decision support alerts.*

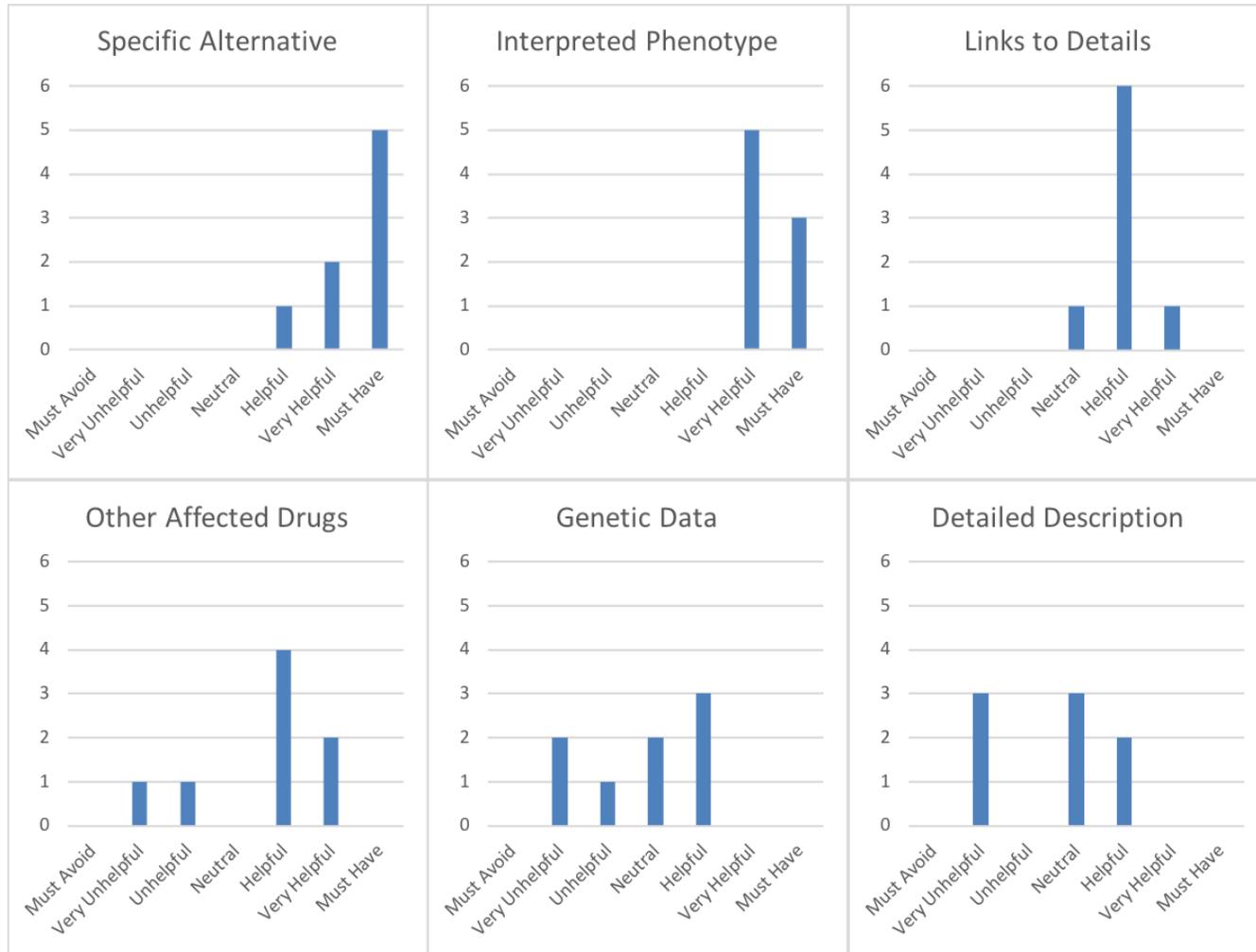

Trusted Knowledge Sources for PGx

When asked where they currently turn for PGx advice, five participants (67.5%) said they use UpToDate, making it the most common response. Only one participant (12.5%) was aware of any currently published PGx prescribing guidelines. Six participants (75%) said they would trust professional societies such as American College of Physicians and American College of Cardiology for PGx recommendations. Five (62.5%) said they would trust a colleague. Only one participant (12.5%) was aware of CPIC prior to the interview, with seven (87.5%) being unfamiliar. Despite this lack of familiarity, seven (87.5%) said they would trust CPIC recommendations once they were told about it. Three participants clarified that they would trust CPIC recommendations, under the assumption that NM had vetted it. Six participants (75%) said links to academic literature published by CPIC embedded into links would increase their trust in an alert's recommendation.

Emergent Themes

Coders reached consensus on five top-level themes with seven underlying axes when analyzing the open-ended responses and elaborations on structured responses. Themes, axes, and common codes are presented in Table 3.

Table 3 – Themes, Axes, and Common Codes from Open-ended Responses

| Themes | Axes | # of Participants Expressing Axis | Most Common Code (# of Participants) |
|---|---|---|---|
| Attitudes and Issues Affecting PGx Use | Discomfort & Skepticism | 8 of 8 | Infrequent PGx User (8) |
| | Education | 6 of 8 | Unlikely to understand underlying genotypic data (2) |
| | | | Information needs reduce with learning effects (2) |
| | Affordability | 6 of 8 | Billing concerns/Insurance coverage (4) |
| Clinical Relevance | N/A | 8 of 8 | Actionable Alerts Only with Clear Guidance (6) |
| Job Flow and System Design | Burden | 8 of 8 | Lack of time (6) |
| | Workflow & Data Display | 8 of 8 | Reviews Medications Earlier (7) |
| | | | Unlikely to visit another screen (4) |
| | Positive Receptivity to Alerts | 7 of 8 | Potentially helpful (4) |
| | Level of Detail | 7 of 8 | Include cost information (5) |
| Trusted Knowledge Sources | N/A | 8 of 8 | UpToDate Professional Societies |
| Patient Communication | N/A | 3 of 8 | No codes appeared more than once |

All eight participants expressed some level of Discomfort & Skepticism in their responses, with the most common concept being that they are Infrequent PGx Users (8 of 8 participants). There was an additional trend in the Discomfort & Skepticism axis with codes that relate to a lack of clarity around the appropriate use of PGx information. The following codes are conceptually related, and taken together, seven of eight participants expressed at least one of them:

- Unsure of how to interpret (4 participants)

- Unsure of how to apply (2 participants)

- Lack of clear guidance (4 participants)

- Lack of clear alternative drugs (2 participants)

- No clear clinical evidence (4 participants)

- Needs clear clinical evidence (3 participants)

- Needs to be demonstrably superior to current standards (2 participants)

- Would use genetics when accepted standard of care (2 participants)

- Not standard of care (1 participant)

These expressions of uncertainty around how to interpret and apply PGx in a clear and standard way, along with the reported infrequent use of PGx information, demonstrates a general discomfort with PGx among the participants.

All eight participants expressed ideas that related to the Clinical Relevance of PGx alerts, with the most common concept being that they are interested in seeing Actionable Alerts Only with Clear Guidance (6 of 8 participants). In other words, they are only interested in seeing alerts that they need to immediately act upon, and which contain a clear recommendation (e.g., *"It's like, give me a clinical recommendation: 'Pick a different statin'" [P03]* and *"If it does not interfere, I do not want to see anything" [P05]*). Other trends in this category include 6 participants expressing that PGx was less helpful for the particular drugs in this study (generally because they do not use those drugs frequently, or they are declining in popularity), 5 participants expressing some form of the idea that they were not the right audience for the alert, 5 participants relating the usefulness of PGx to its timeliness and relevance, and 5 participants expressing that PGx alerts aren't helpful if a patient is already successful on the medication in question.

Seven participants expressed that they review a patient's medication list early in their encounter process. Four stated that they were unlikely to go to a different screen to look up genetic information. Additionally, five participants expressed in some way that lists, such as problem lists and medication lists, are poorly maintained or inconsistent.

All eight participants expressed that incorporating PGx into their practice is, or could become, a burden on them. Six explicitly stated that they had a Lack of Time to think about PGx (e.g. *"In the middle of seeing 20 patients I'm not gonna go looking through that." [P05]*).

Despite the overall negative attitudes reported so far, seven participants expressed a general Positive Receptivity to Alerts. Four explicitly stated that they think genetics-based alerts are potentially helpful (e.g., *"I think that would be helpful if you could get it right" [P03]*) and three expressed generally or conceptually positive attitudes about genetic data being used in alerts.

Analysis of the Level of Detail axis demonstrates that highly detailed alert text is inappropriate. Instead, alerts should be specific in their recommendations without too much detail. Users would be willing to click links to detailed text if they feel they need it. Three participants stated that they Prefer Specificity (e.g. *"I just need very concrete. Like you need to dose 'this' medicine, 'this' often" [P01]*), three stated that Details Are Too Much to Read, and four stated that they Would Use Detailed Links (e.g. *"To actually display that information, that's way too much in an alert. Like a link to it will be fine" [P07]*).

Though participants were never asked about it explicitly, they frequently raised the issue of financial concerns. Drug and genetic testing affordability were frequently cited under Attitudes and Issues Affecting PGx Use. Additionally, five participants expressed a desire to see cost information included in alert text. In particular, they were interested in seeing information about the costs of the recommended alternative drug.

Overall, the themes that emerged during this qualitative analysis paint a clear and coherent narrative. Clinicians are receptive to the idea of using PGx in their practice, but they are currently uncomfortable with PGx and unclear and uneducated on how to use it. They feel a sense of burden about adding it to their workflows. They lack time to think about it, and often find it irrelevant. Therefore, any interventions based on PGx must be highly actionable with clear, concise, and explicitly stated recommendations. Implementers must be selective in which medications they build PGx alerts for and in which clinicians are exposed to them. These results are concordant with the findings from the structured analysis reported above.

**Discussion**

The high level of concordance we found in clinician preferences and attitudes allows us to establish a set of baseline principles upon which to build more effective PGx CDS at NM. Several of the findings are consistent with previous research on non-PGx CDS and the general concepts of being timely and relevant.[25-28] However, others are especially relevant, or even novel, to the use of PGx information. We present these principles in Table 4.

Table 4 – Principles for More Effective PGx CDS Design

| Principle | Description |
|---|---|
| Be Specific and Actionable | Make a clear recommendation to the user with a specific alternative medication and dosage, pre-calculated based off of the patient's genetics. Avoid recommendations to simply "consider alternatives." Do not interrupt the user with informational alerts about a patient's genetic profile if those test results are not immediately relevant. Ensure the user is the right person to see the information. Ensure alerts are built for the right medications. |
| Be Brief | Most physicians are not interested in reading in-depth information on genetics at the point of care, so keep text brief and simple. Make educational materials available through a link for those that want more background. |
| Display Phenotypes not Genotypes | Because most physicians are generally not well trained on genetics, they do not find raw genetic data useful in an alert. Avoid values like "CYP2C19 *2/*2," and instead provide a phenotypic interpretation such as "Clopidogrel Poor Metabolizer." |
| Rely on Sources Clinicians Already Trust | Because genetics-based prescribing is novel to many clinicians, lean on sources they already trust, whenever possible. Use information from UpToDate, established professional societies, or emphasize that your institution has vetted the recommendations coming from unfamiliar sources. |
| Be Adaptable to Learning Effects | As physicians learn and become more comfortable with PGx, their information needs may change over time. Review and revise alert designs on a regular basis or design them to be adaptable in the content they display. |

In Table 5, we compare these principles to prior work in the literature. The Devine, et al., Nishimura, et al., and Melton, et al. works represent prior evaluations of PGx CDS systems that attempted to extrapolate one or more recommendations for alert design from user testing. Nishimura, et al. and Melton, et al. were limited in their attempts to provide broad design recommendations but drew conclusions similar to our "Be Brief" and "Be Specific and Actionable" principles. Our results largely concur with the results of the work by Devine, et al., but with the addition of our "Be Adaptable to Learning Effects" principle. The Horsky, et al. study is a systematic review of traditional (non-PGx) CDS systems that draws from 112 articles.

Of relevance to PGx CDS, they concur with our recommendations around brevity and trust. Their analysis is silent on the topic of genotype and phenotype, as well as on potential learning effects. While they touch on the concepts of specificity and actionability, they draw somewhat different conclusions than us. They recommend CDS that is more "advisory," because "some clinicians may perceive overly prescriptive advice [as] infringing on their sense of professional autonomy." Though we do not recommend PGx CDS alerts be wholly inflexible and commanding in their design, we do find that, likely due to lack of education on the topic, participants in this study preferred stronger, more specific recommendations and showed distaste for more ambiguous or unspecific guidance.

Table 5 – PGx CDS Design Principles in the Literature

*A comparison of the PGx CDS design principles established in this study to design principles established in prior Non-PGx CDS and PGx CDS system evaluations. (Symbol definitions: x = full concurrence between this study and the cited study; * = partial concurrence between this study and the cited study)*

|  | Prior Studies | | | |
|---|---|---|---|---|
|  | Non-PGx | PGx CDS | | |
| **Principle** | **Horsky [26]** | **Devine [29]** | **Nishimura [30]** | **Melton [31]** |
| Be Specific and Actionable | * | x |  | x |
| Be Brief | x | x | x |  |
| Display Phenotypes not Genotypes |  | x |  |  |
| Rely on Sources Clinicians Already Trust | x | x |  |  |
| Be Adaptable to Learning Effects |  |  |  |  |

This comparison begins to shed light on the question of how PGx CDS may be different from other forms of CDS. Clearly, the need for abstraction from genotype to phenotype is unique to the PGx space. Additionally, the need to account for learning effects is likely novel because most other forms of CDS convey information clinicians are likely to have already been formally trained on. Finally, we find evidence in favor of stronger recommendations that rely less on clinician autonomy and are unambiguous about the expected alternative action.

Based on the results of this study, the issue of passive PGx alerting (i.e., optional to view, as opposed to interruptive) deserves particular attention. Of the seven location choices provided, "in an alert on a separate Best Practice Advisories tab in the patient's chart" was the least popular option among clinicians, due to the fact that many users do not regularly check that location in the EHR. This result suggests that passive alerting may not be the most effective approach. However, previous work at NM shows that even infrequent use leads to a significant number of actual alert interactions, due to sheer volume. Passive alerts should not be discounted because their default location is unpopular, but it is a good idea to reconsider where such alerts should appear. Our results suggest that another location that is already part of common workflows may be more appropriate, such as the Problem List or Medications List. In the future, as genetic data use becomes more common, a Genetic Information section may also be appropriate. These solutions will require PGx data standards and EHR support that either do not currently exist or are only in the earliest stages of functionality.

Two particular findings of this study concur with previous PGx literature. The general sentiments of uncertainty and doubt expressed by the clinicians we interviewed are consistent with previous publications documenting the low level of physician education in pharmacogenomics.[3] Additionally, our finding that clinicians hold a generally neutral view of PGx's helpfulness, with some receptivity to its use in the future, is consistent with our previous findings.[23] Given the time difference between when those attitudes were assessed (2015 vs.

2019), it does not appear that those attitudes are changing quickly at NM; nor have those attitudes been altered with further exposure to the first-generation PGx CDS tools at NM.

Our finding that PGx CDS recommendations need to be specific and actionable is at odds with the current state of PGx guidelines.  Currently, the closest the field has to a "gold standard" for PGx CDS is the genetically-based prescribing recommendations published by CPIC.[7] These recommendations are very useful for identifying when a drug-gene interaction is relevant and when to interrupt the clinician, but they do not go so far as to provide a specific recommended alternative.  Therefore, PGx CDS that directly translates CPIC recommendations in their published form is unlikely to be well received by clinicians.  Organizations such as CPIC should strive to provide more specific recommendations in the future, but that will take significant time to achieve, due to the complexities of the decision-making process.  Genetics are only one piece of the equation, with other patient-specific clinical factors being highly relevant. Until the field develops comprehensive algorithms for calculating recommended medications and dosages based off of both genetics and traditional clinical variables, it will likely fall upon individual institutions to create and implement their own rules and recommendations.  This will require substantial institutional commitment and will likely lead to further heterogeneity of practice.

In light of these results, we reiterate the need for standardized PGx interpretations and vocabularies, which we, and others, have called for in previous publications.[13,32,33] Phenotypic interpretations are clearly a critical component of conveying PGx knowledge to clinicians and standardized terms would significantly ease future implementations and EHR support.  Additionally, computable vocabularies for representing this information in EHRs will be necessary for integrating PGx information deeply into the clinician workflow outside of just pop-up alerts (e.g., Problem and Medication List support, and Genetic Information screens).

Finally, in recognition of the fact that most currently-practicing clinicians have not been trained in PGx,[3] we would call upon the sources that clinicians most trust, such as professional societies like the American College of Physicians and American College of Cardiologists, and reference providers like UpToDate, to seriously examine and expand upon the use of PGx data in their guidelines and best practices, when appropriate. Particularly in light of recently issued FDA guidance for CDS, it is critical that there be established clinical guidelines that PGx CDS systems can cite in their recommendations.[34]

**Conclusion**

Through the use of socio-technical principles, this study demonstrates areas in which PGx CDS is different from traditional CDS and provides direction for future PGx CDS designs. Unlike traditional CDS, PGx CDS requires abstraction of clinical data through the use of phenotypes rather than genotypes. Additionally, lack of training means that recommendations must be clear and strong and that there are likely learning effects to account for. Clinicians showed a preference for brief alerts that are clear about the expected action with specific alternative recommended medications and dosages. They also frequently expressed feelings of uncertainty and doubt around the use of PGx, so future designs must focus on simplicity and specificity while referencing well-trusted sources. Designs following these principles are more likely to encourage adherence than those that simply provide information for the clinician to consider on their own. Therefore, current CPIC PGx recommendations do not directly translate to successful CDS because they do not provide specific recommendations. Until specific recommendations become more standardized, any institution interested in using PGx CDS must commit to establishing recommended alternatives. Likewise, the field as a whole must work to develop standardized phenotypic interpretations and vocabularies for clinically relevant PGx results.

**REFERENCES**

1 Gill J, Obley AJ, Prasad V. Direct-to-Consumer Genetic Testing: The Implications of the US FDA's First Marketing Authorization for BRCA Mutation Testing. Jama 2018;**319**(23):2377-78

2 Allyse MA, Robinson DH, Ferber MJ, Sharp RR. Direct-to-Consumer Testing 2.0: Emerging Models of Direct-to-Consumer Genetic Testing. Mayo Clinic proceedings 2018;**93**(1):113-20

3 Stanek EJ, Sanders CL, Taber KA, et al. Adoption of pharmacogenomic testing by US physicians: results of a nationwide survey. Clin Pharmacol Ther 2012;**91**(3):450-8

4 Overby CL, Erwin AL, Abul-Husn NS, et al. Physician Attitudes toward Adopting Genome-Guided Prescribing through Clinical Decision Support. J Pers Med 2014;**4**(1):35-49

5 Phillips KA, Veenstra DL, Oren E, Lee JK, Sadee W. Potential role of pharmacogenomics in reducing adverse drug reactions: a systematic review. Jama 2001;**286**(18):2270-9

6 Weinshilboum R, Wang LW. Pharmacogenomics: Bench to bedside. Nature Reviews Drug Discovery 2004;**3**(9):739-48

7 Caudle KE, Klein TE, Hoffman JM, et al. Incorporation of Pharmacogenomics into Routine Clinical Practice: the Clinical Pharmacogenetics Implementation Consortium (CPIC) Guideline Development Process. Curr Drug Metab 2014

8 Green ED, Guyer MS, National Human Genome Research I. Charting a course for genomic medicine from base pairs to bedside. Nature 2011;**470**(7333):204-13

9 Herr TM, Bielinski SJ, Bottinger E, et al. A conceptual model for translating omic data into clinical action. Journal of pathology informatics 2015;**6**:46 10 Overby CL, Kohane I, Kannry JL, et al. Opportunities for genomic clinical decision support interventions. Genet Med 2013;**15**(10):817-23

11 Rasmussen-Torvik LJ, Stallings SC, Gordon AS, et al. Design and Anticipated Outcomes of the eMERGE-PGx Project: A Multicenter Pilot for Preemptive Pharmacogenomics in Electronic Health Record Systems. Clin Pharmacol Ther 2014;**96**(4):482-89

12 Herr TM, Bielinski SJ, Bottinger E, et al. Practical considerations in genomic decision support: The eMERGE experience. Journal of pathology informatics 2015;**6**:50

13 Herr TM, Peterson JF, Rasmussen LV, Caraballo PJ, Peissig PL, Starren JB. Pharmacogenomic clinical decision support design and multi-site process outcomes analysis in the eMERGE Network. J Am Med Inform Assoc 2019;**26**(2):143-48

14 Hoffman JM, Haidar CE, Wilkinson MR, et al. PG4KDS: a model for the clinical implementation of pre-emptive pharmacogenetics. Am J Med Genet C Semin Med Genet 2014;**166C**(1):45-55

15 O'Donnell PH, Bush A, Spitz J, et al. The 1200 patients project: creating a new medical model system for clinical implementation of pharmacogenomics. Clin Pharmacol Ther 2012;**92**(4):446-9

16 Johnson JA, Elsey AR, Clare-Salzler MJ, Nessl D, Conlon M, Nelson DR. Institutional profile: University of Florida and Shands Hospital Personalized Medicine Program: clinical implementation of pharmacogenetics. Pharmacogenomics 2013;**14**(7):723-6

17 Blagec K, Koopmann R, Crommentuijn-van Rhenen M, et al. Implementing pharmacogenomics decision support across seven European countries: The Ubiquitous Pharmacogenomics (U-PGx) project. J Am Med Inform Assoc 2018;**25**(7):893-98 18 O'Donnell PH, Danahey K, Jacobs M, et al. Adoption of a clinical pharmacogenomics implementation program during outpatient care--initial results of the University of Chicago "1,200 Patients Project". Am J Med Genet C Semin Med Genet 2014;**166C**(1):68-75

19 O'Donnell PH, Wadhwa N, Danahey K, et al. Pharmacogenomics-Based Point-of-Care Clinical Decision Support Significantly Alters Drug Prescribing. Clin Pharmacol Ther 2017;**102**(5):859-69

20 Bell GC, Crews KR, Wilkinson MR, et al. Development and use of active clinical decision support for preemptive pharmacogenomics. J Am Med Inform Assoc 2014;**21**(e1):e93-9

21 Peterson JF, Field JR, Unertl KM, et al. Physician response to implementation of genotype-tailored antiplatelet therapy. Clin Pharmacol Ther 2016;**100**(1):67-74

22 St Sauver JL, Bielinski SJ, Olson JE, et al. Integrating Pharmacogenomics into Clinical Practice: Promise vs Reality. The American journal of medicine 2016;**129**(10):1093-99 e1

23 Herr TM, Smith ME, Starren JB. A Mixed-Methods Analysis of Clinician Response and Adherence to Pharmacogenomic Clinical Decision Support: Northwestern University, 2019.

24 Rasmussen LV, Smith ME, Almaraz F, et al. An ancillary genomics system to support the return of pharmacogenomic results. J Am Med Inform Assoc 2019;**26**(4):306-10

25 Bates DW, Kuperman GJ, Wang S, et al. Ten commandments for effective clinical decision support: making the practice of evidence-based medicine a reality. J Am Med Inform Assoc 2003;**10**(6):523-30 26 Horsky J, Schiff GD, Johnston D, Mercincavage L, Bell D, Middleton B. Interface design principles for usable decision support: a targeted review of best practices for clinical prescribing interventions. J Biomed Inform 2012;**45**(6):1202-16

27 Kawamoto K, Houlihan CA, Balas EA, Lobach DF. Improving clinical practice using clinical decision support systems: a systematic review of trials to identify features critical to success. BMJ 2005;**330**(7494):765

28 Payne TH, Hines LE, Chan RC, et al. Recommendations to improve the usability of drug-drug interaction clinical decision support alerts. J Am Med Inform Assoc 2015;**22**(6):1243-50

29 Devine EB, Lee CJ, Overby CL, et al. Usability evaluation of pharmacogenomics clinical decision support aids and clinical knowledge resources in a computerized provider order entry system: a mixed methods approach. Int J Med Inform 2014;**83**(7):473-83

30 Nishimura AA, Shirts BH, Salama J, Smith JW, Devine B, Tarczy-Hornoch P. Physician perspectives of CYP2C19 and clopidogrel drug-gene interaction active clinical decision support alerts. Int J Med Inform 2016;**86**:117-25

31 Melton BL, Zillich AJ, Saleem J, Russ AL, Tisdale JE, Overholser BR. Iterative Development and Evaluation of a Pharmacogenomic-Guided Clinical Decision Support System for Warfarin Dosing. Appl Clin Inform 2016;**7**(4):1088-106

32 Caudle KE, Dunnenberger HM, Freimuth RR, et al. Standardizing terms for clinical pharmacogenetic test results: consensus terms from the Clinical Pharmacogenetics Implementation Consortium (CPIC). Genet Med 2017;**19**(2):215-23 33 Caudle KE, Keeling NJ, Klein TE, Whirl-Carrillo M, Pratt VM, Hoffman JM. Standardization can accelerate the adoption of pharmacogenomics: current status and the path forward. Pharmacogenomics 2018;**19**(10):847-60

34 FDA. Guidance Document: Clinical Decision Support Software. Secondary Guidance Document: Clinical Decision Support Software  2019. https://www.fda.gov/regulatory-information/search-fda-guidance-documents/clinical-decision-support-software.